\documentclass[review]{elsarticle}

\usepackage{amsmath}
\usepackage{graphicx}
\usepackage{url}
\usepackage{epstopdf}
\usepackage{amsthm}
\usepackage{amssymb}
\usepackage[toc,page]{appendix}
\usepackage{bbold}

\newtheorem*{theorem}{Theorem}

\journal{Journal of \LaTeX\ Templates}

\begin{document}

\begin{frontmatter}

\title{Compressive Sensing of Signals Generated in Plastic Scintillators in a Novel J-PET Instrument}

\author[SWIERK]{L.~Raczy\'nski}
\author[WFAIS]{P.~Moskal}
\author[SWIERK]{P.~Kowalski}
\author[SWIERK]{W.~Wi\'slicki}
\author[WFAIS]{T.~Bednarski}
\author[WFAIS]{P.~Bia\l as}
\author[WFAIS]{E.~Czerwi\'nski}
\author[WFAIS]{A.~Gajos}
\author[WFAIS,PAN]{\L .~Kap\l on}
\author[WCHUJ]{A.~Kochanowski}
\author[WFAIS]{G.~Korcyl} 
\author[WFAIS]{J.~Kowal}
\author[WFAIS]{T.~Kozik}
\author[WFAIS]{W.~Krzemie\'n}
\author[WFAIS]{E.~Kubicz}
\author[WFAIS]{Sz.~Nied\'zwiecki}
\author[WFAIS]{M.~Pa\l ka}
\author[WFAIS]{Z.~Rudy}
\author[WFAIS]{O.~Rundel}
\author[WFAIS]{P.~Salabura}
\author[WFAIS]{N.G.~Sharma}
\author[WFAIS]{M.~Silarski}
\author[WFAIS]{A.~S\l omski} 
\author[WFAIS]{J.~Smyrski}
\author[WFAIS]{A.~Strzelecki}
\author[WFAIS,PAN]{A.~Wieczorek}
\author[WFAIS]{M.~Zieli\'nski}
\author[WFAIS]{N.~Zo\'n}




\address[SWIERK]{\'Swierk Computing Centre, National Centre for Nuclear Research, 05-400 Otwock-\'Swierk, Poland}
\address[WFAIS]{Faculty of Physics, Astronomy and Applied Computer Science,
 Jagiellonian University, 30-059 Cracow, Poland}
\address[PAN]{Institute of Metallurgy and Materials Science of Polish Academy of Sciences, Cracow, Poland.}
\address[WCHUJ]{Faculty of Chemistry, Jagiellonian University, 30-060 Cracow, Poland}

\begin{abstract}
The J-PET scanner, which allows for single bed imaging of the whole human body, is currently under development at the Jagiellonian University. The discussed detector offers improvement of the Time of Flight (TOF) resolution due to the use of fast plastic scintillators and dedicated electronics allowing for sampling in the voltage domain of signals with durations of few nanoseconds. In this paper we show that recovery of the whole signal, based on only a few samples, is possible. In order to do that, we incorporate the training signals into the Tikhonov regularization framework and we perform the Principal Component Analysis decomposition, which is well known for its compaction properties. The method yields a simple closed form analytical solution that does not require iterative processing. Moreover, from the Bayes theory the properties of regularized solution, especially its covariance matrix, may be easily derived. This is the key to introduce and prove the formula for calculations of the signal recovery error. In this paper we show that an average recovery error is approximately inversely proportional to the number of acquired samples.

\end{abstract}

\begin{keyword}
\texttt{Tikhonov regularization} \sep \texttt{Compressed Sensing} \sep \texttt{Positron Emission Tomography} \sep \texttt{J-PET}
\end{keyword}

\end{frontmatter}


\section{Introduction}
Positron Emission Tomography (PET)~\cite{Brownell1953,Robertson1973,Bailey2005} is currently one of the most prominent and promising techniques in the field of medical imaging. It plays a unique role both in medical diagnostics and in monitoring effects of therapy, in particular in oncology, cardiology and neurology. Therefore, notable efforts are devoted to improve this imaging technique. The best way so far is to determine the annihilation point along the Line of Response (LOR) based on measurement of the time difference between the arrival of the gamma quanta at the detectors, referred to as the Time of Flight (TOF) difference~\cite{Karp2008,Kardmas2009}. As it was shown in Ref.~\cite{Moses}, even with the TOF resolution of about 400~ps that is achievable with non-organic crystals, a signal-to-noise ratio can be improved substantially in reconstruction of clinical PET images.

In the articles~\cite{NovelDetectorSystems,StripPETconcept,TOFPETDetector,JPET-Genewa}, a new concept of the TOF-PET scanner, named J-PET, was introduced. The J-PET detector offers improvement of the TOF resolution due to the use of fast plastic scintillators. A single detection unit of the newly proposed TOF-PET detector is built out of a long scintillator strip. Light pulses produced in the strip propagate to its edges where they are converted via photomultipliers into electric signals. There are two main reasons why the TOF resolution may be improved in J-PET scanner: i) a very short rise-time and duration of the signals and ii) a relation between the shape and amplitude of the signals and the hit position. The latter feature usually distorts the time resolution but, when the waveform of the signal is registered, the information about a change of the shape with the position may increase the position resolution and indirectly improve also the resolution of the time 
determination~\cite{ResImpr}. However, to probe the signals, with duration times of few nanoseconds, a sampling time of order of picoseconds is required. This can be done well with the oscilloscopes during the laboratory studies on the prototype, but in the final multimodular devices with hundreds of photomultipliers, probing with oscilloscopes is not feasible~\cite{H_Kim, D_Xi}. Therefore, sampling in the voltage domain using a predefined number of voltage levels is needed. An electronic system for probing these signals in a voltage domain was developed and successfully tested~\cite{palka}.

In recent papers~\cite{NIM14LR, NIM14PM, NIM15PM} we have investigated the performance of a single unit of a J-PET scanner. Sampling in the voltage domain at four thresholds was simulated and each pair of waveforms was represented by a 15-dimensional vector holding information about the relative time values of a signal's arrival at both scintillator ends~\cite{NIM14LR}. In that scenario, the spatial and time resolutions of the hit position and event time for annihilation quanta were determined to be 1.05 cm and 80 ps ($\sigma$), respectively. It is evident that the spatial and time resolutions can be further improved primarily by an increase in the number of threshold levels, as was also concluded  e.g. in article~\cite{ResImpr}. However, the number of channels in the electronic devices is a very important factor in determining the cost of the PET scanner. Therefore, the question arises: is it possible to recover the whole signal based on only a few samples? Equivalently, how many threshold levels have to be applied to achieve a reasonable estimation error?

In this article we propose a novel signal recovery scheme based on ideas from the Tikhonov regularization~\cite{Tik1, Tik2} and Compressive Sensing~\cite{CStheory1,CStheory2} methods that is compatible with the signal processing scenario in J-PET devices. We investigate the quality of signal recovery based on the scheme with a single scintillator strip module introduced in Ref.~\cite{NIM14LR, NIM14PM}. The two most important aspects of our work involve i) a development of fast recovery algorithms and ii) a statistical analysis of an error level. In practice the algorithm needs to work in real-time scenarios: during a single PET examination more than 10 million signals are acquired in just 10-15 minutes. Moreover, only results for realistic scenarios with noisy measurements are considered. In particular, as was mentioned, the most important part of our investigations is to determine a dependence of the signal recovery error on the number of samples taken in the voltage domain. In this paper the formula for calculations of the recovery error will be introduced and proven.   

This article is organized as follows. We will define the problem of signal recovery and show briefly the Tikhonov regularization and compressive sensing methods in Sec.~2. In the last part of this section we will introduce the theorem enabling the determination of the signal recovery error as a function of the number of samples. The experimental setup of the simplified PET device with a single scintillator strip that enables us to acquire real signals as well as the results of their analysis are presented in Sec.~3. A detailed analysis of the experimental characteristic of signal recovery error as a function of the number of samples, as well as the explanation of the specificity of the signal recovery method in the application to the J-PET measurement are provided in Sec.~3.2. In Sec.~3.3 we have discussed the limitations of the method of signal recovery. In particular, we have presented how the quality of the information needed to recover the signals, and therefore to estimate the recovery error, vary with size of training set of fully acquired signals. Moreover, we have demonstrated that using the recovered waveform of the signals, instead of samples at four voltage levels alone, improves the spatial resolution of the hit position reconstruction. A detailed description of this study is given in Sec.~3.4. The conclusions and directions for future work are presented in Sec.~4.

\section{Materials and methods}

\subsection{Problem definition}

We wish to recover a finite signal $y^0 \in \mathcal{R}^N$ in a situation where the number $M$ of available samples, denoted as measurement $y \in \mathcal{R}^M$, is much smaller than the signal dimension $N$ ($y$ is sampled on some partial subset $\Omega$, where the cardinality $|\Omega|=M$). In the compressive sensing (CS) method~\cite{CStheory1,CStheory2}, a sparse expansion 
$x^0 \in \mathcal{R}^N$ of signal $y^0$, evaluated via linear and orthonormal transformation $y^0=A x^0$, is considered. In the following we assume we are given a contaminated measurement $y$ and then one may write: 
$y = A_{\Omega} x^0 +e;$ where $A_{\Omega}$ is a $M \times N$ matrix modeling the sampling system, constructed from $M$ rows of matrix $A$ that corresponds to the indexes of $y$ described in the subset $\Omega$, and $e$ is an error term. Therefore, during the recovery process the information about the measurement $y$ may be included in the form of the linear system of equations:
\begin{equation}
	y = A_{\Omega} x.
        \label{LinearSyst}
\end{equation}
It should be stressed that in the case of presence of noise, represented by signal $e$, instead of an exact recovery of signal $x^0$ we will consider the solution $\hat{x}$ and by the analogy, instead of signal $y^0$ we will consider the solution 
$\hat{y}$. Before we look in more detail we may state that the evaluation of $\hat{y}$ requires two steps: i) recovery of the sparse expansion $\hat{x}$ and ii) calculation of $\hat{y}$ based on the 
$\hat{x}$. The first step of the procedure is crucial. 

In the situation where an exact solution cannot be found, CS method provide an attempt to recover $\hat{x}$ by solving optimization problem of the form
\begin{equation}
	\hat{x} = \arg \min ||x||_1 ~~~~ \textnormal{such that} ~~~~ ||y - A_{\Omega} x||_2 \le \epsilon \nonumber
\end{equation}
where $\epsilon$ is the size of the error term $e$. The $l_1$ minimization approach provides a powerful framework for recovering sparse signals. Moreover, the use of $l_1$ minimization leads to a convex optimization problems for which there exist a variety of greedy approaches like Orthogonal Matching Pursuit~\cite{CSomp} or Basis Pursuit~\cite{CSbp}. Other insights provided by CS are related to the construction of measurement matrices ($A_{\Omega}$) that satisfy the Restricted Isometry 
Property~\cite{CSrip1,CSrip2}. For an extensive review of CS the reader is referred to 
Refs.~\cite{CStheory1,CStheory2,CSrip1,CSrip2}.

We will incorporate from the CS framework to our scheme only the idea of conducting the experiment, formulated by a linear system of equations as given explicitly in Eq.~(\ref{LinearSyst}). The problem formulated by Eq.~(\ref{LinearSyst}) alone is essentially underdetermined, and is so-called ill-posed~\cite{Had23}. As in the case of the CS method, it is necessary to incorporate further assumptions or information about the desired solution in order to stabilize the problem. As an alternative to the CS theory, one may use the regularization methods~\cite{Tik1, Tik2, TSVD1, TSVD2}. The Tikhonov regularization (TR) method~\cite{Tik1, Tik2} is the most suitable for our problem. Here, the idea is to define the regularized solution $\hat{x}$ as the minimizer of the following expression:
\begin{equation}
	\hat{x} = \arg \min \{(y-A_{\Omega}x)^{T} R^{-1} (y-A_{\Omega}x)  + (x - \mu)^{T} P^{-1} (x - \mu) \}.
        \label{TikMin}
\end{equation}
In Eq.~(\ref{TikMin}), both signals $y$ and $x$ are assumed to be given with multivariate normal (MVN) distributions,
\begin{align}
	& y \sim \mathcal{N}(A_{\Omega}x, R),   \label{Ydist}    	\\
	& x \sim \mathcal{N}(\mu, P),      		\label{Xdist}    	
\end{align}
where $A_{\Omega}x$ and $R$ are the mean value and covariance matrix of a measured signal $y$, respectively, $\mu$ and $P$ are the mean value and covariance matrix of a prior distribution of $x^0$, respectively. The covariance matrix $R$ in Eq.~(\ref{Ydist}) is diagonal with the values on the diagonal equal to the measurement error variances $\sigma^2$ (as explained in Sec.~2.4). With the introduction of the second term to the optimization problem in Eq.~(\ref{TikMin}), an additional information from a training set of fully sampled signals $y^0$ is provided. The prior distribution of sparse representation $x^0$ (see Eq.~(\ref{Xdist})) is evaluated based on the linear transformation of the training set of signals $y^0$ by using the Principal Component Analysis (PCA) decomposition~\cite{PCA}. Thus, in order to find the sparse representation $\hat{x}$ of a given measurement $y$, as a solution of Eq.~(\ref{TikMin}), one needs to specify first the prior distribution of $x^0$. 

Beside the advantage of including the additional information from training signals, a further benefit of the TR approach is that the problem in Eq. (2) has an optimal solution which can be determined explicitly. In Sec.~2.2 we will evaluate the orthonormal matrix $A$, as well as the parameters of the prior distribution of signal $x^0$, see Eq.~(\ref{Xdist}), via the PCA decomposition of training signals $y^0$. It should be stressed that these parameters are calculated only once, at the preparation stage of the procedure. Thus, the same matrix $A, P$ and vector $\mu$, are used to recover a signal $\hat{x}$ for each measurement $y$. An example of the idea of using the PCA decomposition of a training data set in a similar problem may be found in Ref.~\cite{Mahal}. In Sec.~2.3 the solution of the TR formula described in Eq.~(\ref{TikMin}), as well as its properties, will be provided. Finally in Sec.~2.4 we will introduce the theorem enabling the determination of the signal recovery error as a function of the number of samples.

\subsection{Principal Component Analysis}

PCA is a statistical study, based on the orthogonal transformation, to convert a set of signals into a set of linearly independent variables, such that the variance of the projected data is maximized. For the training data matrix of fully sampled signals
\begin{equation}
	Y = [ (y^0_{(1)}-m) | (y^0_{(2)}-m) | ... | (y^0_{(L)}-m)],
        \label{dataY}
\end{equation}
where $m$ is the mean of the aligned $L$ training signals $y^0$, the PCA coordinates 
$X = [x^0_{(1)} | x^0_{(2)} | ... | x^0_{(L)}] $ are given by:
\begin{equation}
	X = A^T Y.
        \label{XAY}
\end{equation}
The matrix $A = [a_{(1)} | a_{(2)} | ... | a_{(N)}] $ in Eq.~(\ref{XAY}) is calculated in such a way that the projection of the data matrix $Y$ with successive basis vectors $a_{(1)}, a_{(2)}, ... , a_{(N)}$ inherits the greatest possible variance in the data set $Y$. Thus, the first basis vector has to satisfy:
\begin{equation}
	a_{(1)} = \arg \max || a^T Y ||_2^{2}, \nonumber
\end{equation}
where $||a||_2=1$ (the orthonormality is restricted). The $k^{\text{th}}$ component can be found by subtracting the first 
$k - 1$ principal components from data set $Y$:
\begin{equation}
	Y_{k} = Y - \sum_{l=1}^{k-1} a_{(l)} a_{(l)}^T Y , \nonumber
\end{equation}
and then finding the basis vector which extracts the maximum variance from this new data matrix
\begin{equation}
	a_{(k)} = \arg \max || a^T Y_{k} ||_2^{2}, \nonumber
\end{equation}
where $||a||_2=1$. 

In the case discussed in this paper, the matrix $A$ is evaluated based on the PCA decomposition of the training set of signals $y^0$ and therefore the parameters of the MVN distribution of $x^0$ ($\mu, P$) are estimated based on data matrix $X$, constructed according to Eq.~(\ref{XAY}). The empirical covariance matrix $P$ of data set $X$ may be evaluated as:
\begin{equation}
	P=E[X \cdot X^T].
	\label{covP}
\end{equation}
The covariance matrix $P$ is diagonal, with values sorted in non-increasing order. Since the mean of the signals in data set $Y$ is equal to 0, see Eq.~(\ref{dataY}), the mean $\mu = 0$.

\subsection{Tikhonov regularization}

In the previous section we have shown how the prior information from a training set of signals $y^0$ i.e. the orthonormal matrix $A$, and the parameters of the prior distribution of signal $x^0$, may be introduced to the TR framework. In this section we will derive a sparse solution $\hat{x}$ of Eq.~(\ref{TikMin}), and its covariance matrix, denoted hereafter as $S$, for a particular measurement $y$, based on the TR assumptions~\cite{Tik1, Tik2}. The posterior probability density function (pdf) of the signal $x$ conditional on measurement $y$, namely $p(x|y)$, can be computed after combining the prior distribution of $x$, $p(x)$, likelihood of measurement $p(y|x)$, and $p(y)$ via the well-known Bayesian rule:
\begin{equation}
	p(x|y) = \frac{p(x) \cdot p(y|x)}{p(y)}.
	\label{Bayes}
\end{equation}
To describe the MVN distribution in Eq.~(\ref{Bayes}) we will use the following notion:
\begin{equation}
	\mathcal{N}(z |u, Q) = \frac{1}{(2 \pi)^{N/2} |Q|^{1/2} } \exp( - \frac {1}{2} (z-u)^T Q^{-1} (z-u)) \nonumber
\end{equation}
where $z$ is an $N-$dimensional variable with mean value $u$ and covariance matrix $Q$. Hence, the marginal and conditional densities of $x$ and $y$ from Eq.~(\ref{Bayes}) are given as follows:
\begin{align}
	p(x) &= \mathcal{N}(x|\mu, P),   \label{probX}       \\   
	p(y|x) &= \mathcal{N}(y|A_{\Omega}x, R), \label{probYX} \\
	p(y) &= \alpha, 				\label{probY}	\\
	p(x|y) &= \mathcal{N}(x| \hat{x}, S).   \label{probXY}    
\end{align}
Equations~(\ref{probX}) and~(\ref{probYX}) result directly from the previously described Eq.~(\ref{Xdist}) and~(\ref{Ydist}), respectively. Equation~(\ref{probY}) shows that the probability $p(y)$ is independent of $x$, and therefore serves as a normalization constant. The posterior probability in Eq.~(\ref{probXY}) can be described exclusively by its first two moments 
($\hat{x}, S$) because a Gaussian pdf is self-conjugate and the pdfs on the right hand side of Eq.~(\ref{Bayes}) are Gaussian. After some simple calculations the equations for $\hat{x}$ and 
$S$ are given by~\cite{Hans97}:
\begin{align}
	\hat{x} &= (P^{-1} \mu + A_{\Omega}^T R^{-1} y) \cdot (P^{-1} + A_{\Omega}^T R^{-1} A_{\Omega})^{-1}, \label{estX} \\
      S &= (P^{-1} + A_{\Omega}^T R^{-1} A_{\Omega})^{-1}. \label{estS}
\end{align}
It is worth noting that the solutions in Eq.~(\ref{estX}) and (\ref{estS}) are analogous to Kalman filter update 
equations (cf. Refs.~\cite{Kal60, Sorensen70}). It can be easily shown that $\hat{x}$ is not only the minimum mean square error (MSE) estimator (see 
Eq.~(\ref{TikMin})) but also the maximum a posteriori (MAP) estimator, that is
\begin{equation}
	\hat{x} = \arg \max \{ p(x|y) \}. \nonumber
\end{equation}

It should be stressed that all the information from the training set of signals $y^0$ (matrix $A, P$ and vector 
$\mu$) and from the oscilloscope specification (matrix $R$) are evaluated only once, at the preparation stage. Thus, the sparse signal $\hat{x}$ may be found, according to Eq.~(\ref{estX}), as a linear combination of the previously defined parameters and a given measurement $y$. However, the evaluation of the covariance matrix $S$, according to Eq.~(\ref{estS}), does not require the information about the measurement $y$, and may be provided at the preparation stage. This fact opens a possibility for an estimation of the theoretical value of the recovery error. This idea will be presented in the next section.   

\subsection{Analysis of the signal recovery error}

As mentioned at the beginning of the Sec.~2, the evaluation of the recovered signal $\hat{y}$ requires two steps: i) recovery of the compact representation $\hat{x}$ via Eq.~(\ref{estX}) and  ii) calculation of $\hat{y}$ as the solution
\begin{equation}
	\hat{y} = A \hat{x} + m,
	\label{estY}
\end{equation}
where $m$ and $A$ are derived by PCA decomposition. One of the benefit of using the TR approach is that it provides an easy way to obtain the error term of the recovered signal $\hat{y}$. We assume for the sake of simplicity that $m$ in Eq.~(\ref{estY}) is known exactly. Since the matrix $A$ is orthonormal, we have $||\hat{y}-y^0||_2=||\hat{x}-x^0||_2 $, and therefore we may focus on the recovered signal $\hat{x}$ error.

In multivariate statistics, the trace of the covariance matrix is considered as the total variance. We will denote the trace of covariance matrix $S$ as $\sigma_x^2$. It is worth noting that $\sigma_x^2$ is the mean value of the recovery error squared norm $||\hat{x}-x^0||_2^2 $. Let $P(k)$ be the $k^{\text{th}}$ diagonal element of covariance matrix $P$ (see Eq.~(\ref{covP})). Find the smallest value $D$, and largest value $\tau$ (with constraints $D>0$ and $\tau>0$) such that for each $1 \le k \le N$:
\begin{equation}
	P(k) \le D \cdot e^{-\tau k}.
	\label{Pk}
\end{equation}
From Eq.~(\ref{Pk}) one may see that $\tau$ controls the decrease rate of $P(k)$: the greater $\tau$, the faster the decreasing of $P(k)$ and better the compressibility of signal $x$. The characteristics $D$ and $\tau$ of the prior distribution of signal $x$ and a standard deviation of noise ($\sigma$) enable us to provide the formula for average value of the recovery error 
$\sigma_x^2$. For this purpose we formulate the following theorem:

\begin{theorem}
Suppose that $D$ and $\tau$ describe the decrease rate of variances of signal $x$ according to Eq.~(\ref{Pk}). The signal $x$ may be recovered as the solution to Eq.~(\ref{estX}) with an average value of error  
\begin{equation}
	\sigma_x^2 \approx \frac{\sigma^2 N}{M \tau} \cdot \log \left( \frac{\sigma^2 N + MD}{\sigma^2 N} \right) . 
	\label{sigmaX}
\end{equation}
\end{theorem}
Equation~(\ref{sigmaX}) enables us to estimate the number of required samples $M$ of signal to achieve a preselected mean recovery error. Intuitively, the $\sigma_x^2$ is also closely related to the compressibility of signal $x$, and from Eq.~(\ref{sigmaX}) one may observe that an average recovery error is inversely proportional to the constant value $\tau$. The proof of the theorem is given in the Appendix.

\section{Experimental results}

\subsection{Experimental setup}

In this section, we present results illustrating the proposed approach and demonstrating that the number of samples ($M$) required to sense the data can be considerably less than the total number of time samples ($N$) in the reference signal $y^0$. 
We investigate the performance of the algorithm using a data set of reference signals registered in single module scintillator strip EJ-230~\cite{EJ} of J-PET device~\cite{NIM14PM}.

\begin{figure}[h!]
	\centerline{\includegraphics[width=0.7\textwidth]{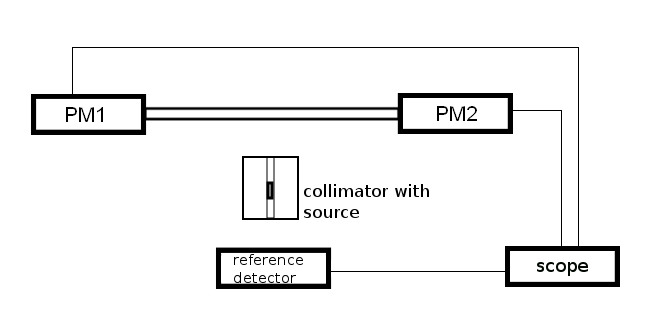}}
\caption{
Scheme of the experimental setup. 
\label{Exper_setup}
}
\end{figure}

The scheme of the experimental setup is presented in Fig.~\ref{Exper_setup}. The 30 cm long strip was connected on two sides to the R4998 Hamamatsu~\cite{HAMAMATSU} photomultipliers denoted as PM1(2). A series of measurements was performed using collimated gamma quanta from a $^{22}$Na source placed between the scintillator strip and the reference detector. The collimator was located on a dedicated mechanical platform allowing it to be shifted along the line parallel to the scintillator strip with a submillimeter precision.
The $^{22}$Na source was moved from the first to the second end in steps of 6~mm. At each position, about 5~000 pairs of signals from PM1 and PM2 were registered in coincidence. These signals were sampled using the Serial Data Analyzer (Lecroy SDA6000A) with a probing interval of 50~ps. To demonstrate the recovery performance only signals from PM1 were investigated (the procedure with signals from PM2 would be the same). The length of a signal $y^0$ was set to 15~ns, which corresponds to $N=300$ samples (see Fig.~\ref{Examp_008}-\ref{Examp_082} for details).

We wish to make one comment about the data acquisition. The signal captured by an oscilloscope is length $N$, where each sample is contaminated with white noise with 0 mean and $\sigma^2$ variance. The simulation of measurement $y$ is then based on selecting $M$ samples according to the subset $\Omega$. However, in order to extract the reference, noise-free signal $y^0$, the acquired $N$ samples have to be subjected to low pass filtering. In the following procedure we will need the signals $y$ and $y^0$ as well.
	
Since the absolute registration time has no physical meaning, we synchronize the signals in data set $Y$ in such way that the fixed index number 20 corresponds to the amplitude of -0.06~V on the rising slope of each signal (see Fig.~\ref{Examp_008}, ~\ref{Examp_026}, ~\ref{Examp_082}). The complete data set $Y$ contains more than 200~000 signal examples and was divided into two disjoint subsets: training and testing part, with a ratio 9 to 1, respectively. In the training data set only the signals $y^0$ are stored, while in the testing one both signals $y$ and $y^0$ are required. 

\subsection{Error recovery investigations}

The training data set $Y$ was transformed via PCA into a new space $X$ according to the scheme shown in Sec. 2.2. The evaluated matrix $A$, as well as the mean value signal $m$, were saved and used in the further analysis during the signal $x$ recovery from the testing data set. In order to find the theoretical value of mean recovery error $\sigma_x^2$, introduced in Eq.~(\ref{sigmaX}), one needs to specify additionally the following parameters: $\sigma, D, \tau$ (we will investigate the 
error $\sigma_x^2$ as a function of the number of samples $M$). The standard deviation of the noise ($\sigma$) was estimated based on the training data set $Y$ to c.a. 0.015~V, which is consistent with the oscilloscope specification. The unknown parameters $D, \tau$ were found after the analysis of diagonal elements of the covariance matrix $P$ of the training data set 
$X$. The smallest value $D$ and the largest value $\tau$ for which the condition from Eq.~(\ref{Pk}) was met, are equal to 4.2~V$^2$ and 0.33, respectively.

\begin{figure}[h!]
	\centerline{\includegraphics[width=0.7\textwidth]{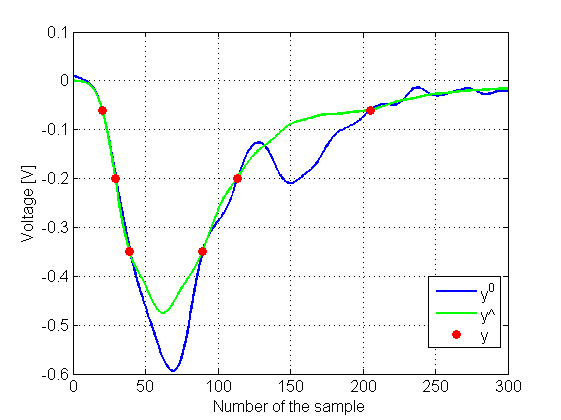}}
\caption{
Signal recovery example: the recovery error is about 0.814 V$^2$.
\label{Examp_082}
}
\end{figure}

It should be stressed that, for a given number of samples ($M$), the expected value of $\sigma_x^2$ in J-PET scenario would be slightly greater than for the one described by Eq.~(\ref{sigmaX}). The reason is that in the J-PET scenario the signals are probed in the voltage domain and hence in the case when the amplitude of the signal is smaller than the threshold level, not all the samples of the signal are acquired (see Fig.~\ref{Examp_082} for example). Therefore, in order to evaluate the theoretical function of mean recovery error in the J-PET scenario, both, the values of the threshold levels as well as the distribution of signal amplitides have to be specified first. 

\begin{figure}[h!]
	\centerline{\includegraphics[width=0.7\textwidth]{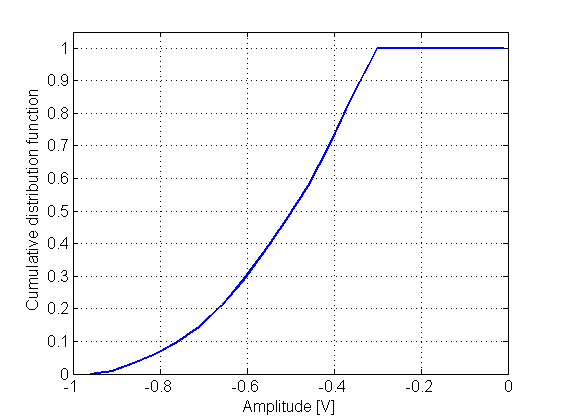}}
\caption{
Experimental cumulative distribution function of signal amplitides.
\label{CDF}
}
\end{figure}

In the first step of the analysis the distribution of signal amplitudes was investigated. The experimental cumulative distribution function (cdf), based on the signals registered at all the positions along the scintillator strip, is presented in 
Fig.~\ref{CDF}. The amplitudes of the signals are in the range from -0.3~V to -1.0~V.
In order to suppress events when gamma quanta were scattered inside the patient’s body, in the current PET scanners (detecting gamma quanta based on photoelectric effect) the energy window, typically in the range from 350~keV to 650~keV, is applied~\cite{Humm}. Such window suppress scattering under angles larger than 60 degrees. The J-PET detector is made of plastic scintillators which are composed of carbon and hydrogen. Due to the low atomic number of these elements the interaction of gamma quanta with energy of 511~keV is predominantly due to the Compton effect whereas the interaction via photoelectric effect is negligible. In order to suppress scattering in the patient through angles larger than 60 degrees, in the J-PET scanner only the signals with energy deposition larger than 200~keV will be accepted~\cite{TOFPETDetector}. Therefore, the signals with amplitude smaller than a -0.3~V are filtered out, and a sharp edge of the spectrum for this value is seen in Fig.~\ref{CDF}.

\begin{figure}[h!]
	\centerline{\includegraphics[width=0.7\textwidth]{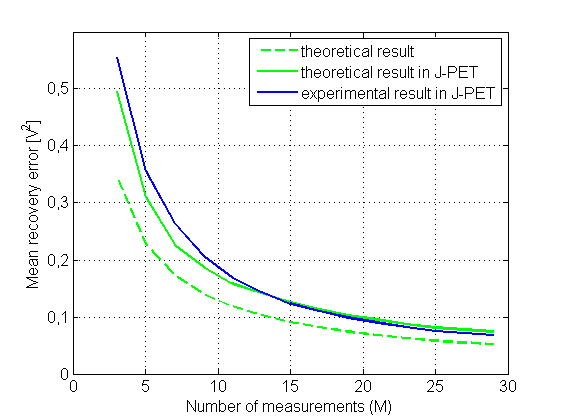}}
\caption{
Comparison of average recovery errors $\sigma_x^2$ as a function of the acquired samples ($M$). Meaning of the curves is described in the text.
\label{Theor_error}
}
\end{figure}

In the next step, based on the fully sampled signals stored in testing data set $Y$, we simulate a front-end electronic device that probes the signals at preselected number of voltage levels, both on the rising and falling slope. We carried out the experiments for different numbers of voltage levels from 2 to 15. In each case, the level of -0.06~V on the rising slope was applied for triggering purposes, as was mentioned in Sec.~2.1. The remaining amplitude levels were adjusted after a simple optimization process, where the goal was to minimize the experimental mean recovery error $\sigma_x^2$. At each step of the optimization process, for a fixed number ($M$) and values of voltage levels, signal recovery was conducted in the following way. For each 300 signal samples from testing data set $Y$, all samples at preselected voltage levels were selected to simulate the measurement $y$. Since the amplitude of the signal may be less than certain voltage levels, not all samples had to be registered. Therefore, for each processed signal, the number of acquired samples would be smaller or equal to $M$. In order to remove the mean value from the measurement $y$, the corresponding values of signal $m$ were subtracted from signal samples from the oscilloscope. The measurement matrix $A_{\Omega}$ was formed from the proper rows of matrix $A$. The signal $\hat{x}$ was recovered using Eq.~(\ref{estX}), and finally the signal $\hat{y}$ was derived as the linear solution of Eq.~(\ref{estY}). With optimized values of voltage levels, theoretical and experimental curves describing the mean recovery error $\sigma_x^2$ as a function of the number of samples ($M$) in the J-PET scenario are evaluated and shown in Fig.~\ref{Theor_error}.

An empirical mean value of $\sigma_x^2$ is marked with a solid blue line in Fig.~\ref{Theor_error} and is very similar to the expected, theoretical characteristic that takes into account the distribution of amplitudes and optimized values of voltage levels (solid green line). The difference between those two functions is larger for small values of $M$ (about $10\%$ of 
$\sigma_x^2$) and almost negligible for greater numbers of samples. However, both of these functions differ significantly from the theoretical characteristic of $\sigma_x^2$, calculated according to Eq.(~\ref{sigmaX}), marked with dashed green line in 
Fig.~\ref{Theor_error}. In the following we will investigate only the case with a four-level measurement, which is of most importance since the currently developed front-end electronic allows one to probe the signals at four fixed-voltage levels. It is evident that this comparison of results may be performed in the same way for all values of $M$.    

The optimized values of the four voltage levels are: -0.06, -0.20, -0.35 and -0.60~V. Since, the index of the sample taken at the voltage level of -0.06~V at the rising slope is common for all signals, the effective number of simulated samples at rising and falling edge is equal to $M=7$. From Fig.~\ref{Theor_error}, the theoretical value of the average recovery error $\sigma_x^2$ for $M=7$ is c.a. 0.173~V$^2$ (dashed green line). However, based on the experimental distribution of amplitudes of the signals, presented in Fig.~\ref{CDF}, only for about $30\%$ of signals would all samples from four thresholds be available (amplitudes larger than -0.60~V). Moreover, for signals with amplitudes in the range from -0.35 V to -0.60 V (about $55\%$ of signals), the effective number of samples is equal to 5 and the theoretical value of $\sigma_x^2$ increases to 0.228~V$^2$. For the rest of the considered signals, with amplitudes in the range from -0.30 V to -0.35 V (about $15\%$ of signals), the effective number of samples is equal to 3 and the theoretical value of $\sigma_x^2$ is 0.346~V$^2$. Finally, the expected mean value of $\sigma_x^2$ in the J-PET scenario for four voltage levels is equal to c.a. 0.227~V$^2$ and is much more comparable with the experimental value (equal to c.a. 0.264 V$^2$) than the theoretical value for 7 samples.    

The analysis of the characteristic of $\sigma_x^2$ allows us to indicate the proper number of samples needed. The function 
$\sigma_x^2(M)$ is approximately proportional to $1/M$ but, due to the logaritmic factor (see Eq.(~\ref{sigmaX})), it drops rapidly until $M$ reaches the value of about 10. Further increase in the number of samples does not provide any significant improvement in the signal recovery. This is very important information since the currently developed front-end electronic enable one to probe the signals at four fixed-voltage levels, providing eight time values for each signal. 

\begin{figure}[h!]
	\centerline{\includegraphics[width=0.7\textwidth]{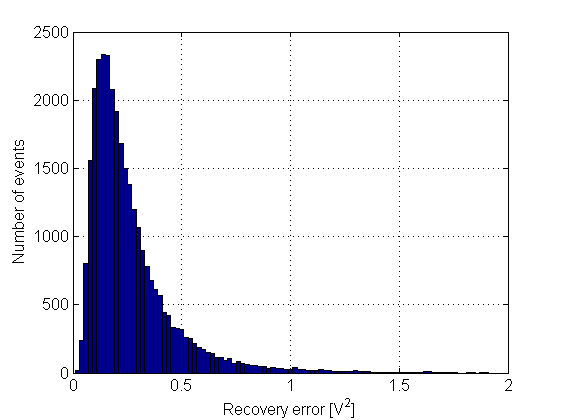}}
\caption{
Distribution of the recovery error evaluated using signals from the testing data set.
\label{Dist_error}
}
\end{figure}

\begin{figure}[h!]
	\centerline{\includegraphics[width=0.7\textwidth]{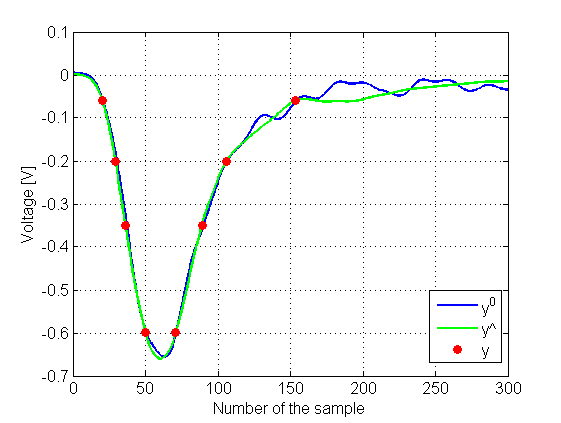}}
\caption{
Signal recovery example: the recovery error is about 0.082 V$^2$.
\label{Examp_008}
}
\end{figure}

\begin{figure}[h!]
	\centerline{\includegraphics[width=0.7\textwidth]{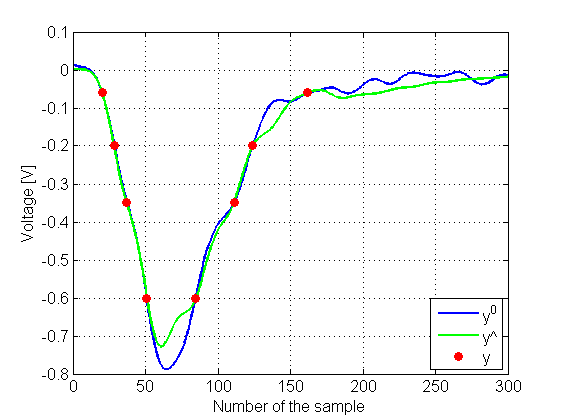}}
\caption{
Signal recovery example: the recovery error is about 0.266 V$^2$.
\label{Examp_026}
}
\end{figure}

The distribution of the recovery error $||x^0 - \hat{x}||_2^2$ evaluated using all signals from the testing data set for optimized values of four voltage levels is shown in Fig.~\ref{Dist_error}.
From the empirical characteristics of $||x^0 - \hat{x}||_2^2$ one may see that the recovery error is concentrated between 0 and 
0.4~V$^2$ with the tail reaching the value 1.5~V$^2$. As was shown in Fig.~\ref{Theor_error}, the mean value in the experiment is equal to c.a.~0.264~V$^2$. In addition, the standard deviation and the median of a probability distribution of a recovery error are equal to c.a.~0.192~V$^2$ and 0.206~V$^2$, respectively. The three signal recovery examples, with small, medium and large recovery error, are shown in Fig.~\ref{Examp_008},~\ref{Examp_026} and~\ref{Examp_082}, respectively.

The values of the signal recovery errors in Fig.~\ref{Examp_008} to~\ref{Examp_082} are as follow: 0.082, 0.266, 0.814~V$^2$. As expected, the worst situation takes a place when the amplitude of the signal is slightly below the selected threshold level (see Fig.~\ref{Examp_082}) or where it is much larger than the highest sampling voltage. In our sampling scheme the highest recovery error occurs for signal amplitudes in the range from -0.55 to -0.6~V and from -0.95 to -1~V (where -1~V corresponds to the maximum amplitude, see Fig.~\ref{CDF}). Unfortunately, there is no possibility to overcome these phenomena when only a few samples of the signal are measured. On the other hand, it can be seen that the mean value of the error $||x^0 - \hat{x}||_2^2$ is on an acceptable level. In a typical situation the signal is recovered quite accurately (see Fig.~\ref{Examp_026}). 

\subsection{Method limitations}

Although, the experimental and theoretical functions describing the recovery errors in the J-PET scenario, presented in 
Fig.~\ref{Theor_error}, are largely consistent, there are at least two aspects of the method that need to be investigated: 

1) the assumption about the prior MVN distribution of signals $x^0$ (see Eq.~(\ref{Xdist})), which has an impact on the difference bewteen the values of the $\sigma_x^2$ errors, 

2) the evaluation of the empirical values of $\sigma_x^2$ as a function of the size of training set of signals $y^0$.

In order to verify the assumption about the normality of signals $x^0$, we have used the Kolmogorov-Smirnov test on each of 
$N$ principal components in the training data set $X$, evaluated according to Eq.~(\ref{XAY}). In each dimension, the mean value as well as the standard deviation were estimated for all the samples. The significance level used in this study was 0.05. The hypothesis, regarding the normal distribution form, was rejected only for the first principal component that holds about $40\%$ of the signal energy. However, in that case the calculated value from the statistical test was not significantly higher than the critical value. From this analysis one infers that the signals stored in matrix $X$ are not exactly normally distributed. This fact may contribute to the difference between the theoretical and empirical values of $\sigma_x^2$.
	
\begin{figure}[h!]
	\centerline{\includegraphics[width=0.7\textwidth]{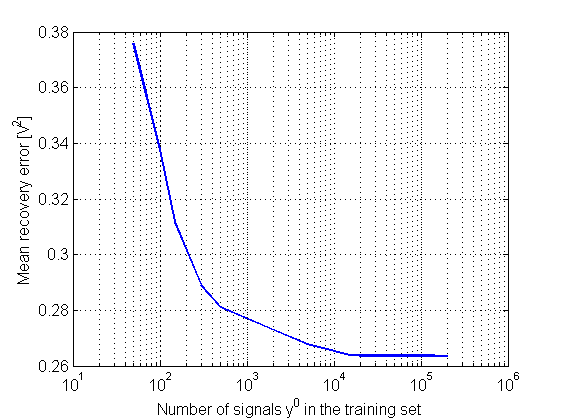}}
\caption{
Influence of the size of the training set of signals $y^0$ on the average recovery error $\sigma_x^2$.
\label{TrainingSet}
}
\end{figure}

It should be stressed that all the informations needed to recover the signal $\hat{x}$ (matrix $A, P$ and vector $\mu$ - see Sec.~2.2 for details), and therefore the empirical values of $\sigma_x^2$, were evaluated based on large set of about 200 000 training signals $y^0$. In the following we will analyze the influence of the size of the training set of signals on the value of the signal recovery error. We conduct the experiment for a wide range of number of signals $y^0$ in the training set from 50 to 200~000. In each case we investigate only the four-level measurement $y$ ($M=7$). The results of the analysis of the empirical values of $\sigma_x^2$ as a function of the size of the training set of signals $y^0$ are shown in Fig.~(\ref{TrainingSet}). 

The parameters of the model that were used for the signal recovery in the study described in Sec.~3.2 were based on 200~000 training signals, which corresponds to the value of the empirical recovery error of 0.264~V$^2$. From Fig.~(\ref{TrainingSet}) one may observe that reducing the number of training signals down to about 10~000 does not influence the quality of the recovery of signal $\hat{x}$; the $\sigma_x^2$ error is almost constant in that range. However, for smaller number of traning signals, the $\sigma_x^2$ error increases rapidly and the recovery of the signal $\hat{x}$ becomes increasingly less accurate. 

\subsection{Spatial resolution of the hit-position reconstruction}

In this section, we will incorporate the method for hit-position reconstruction, described in Ref.~\cite{NIM14LR}, in order to evaluate a position resolution of the J-PET scanner with fully recovered signals. We will compare the spatial resolutions obtained from the original raw-signal (300 samples) to those from the compressed signal (e.g. 8 samples). We have carried out experiments with numbers of voltage levels from 2 to 15, which corresponds to the number of samples $M$ from 3 to 29. 

For a single event of gamma quantum interaction along the scintillator strip, a pair of signals at two photomultipliers is measured in a voltage domain. Next, the signals are recovered according to the description in Sec.~2, and finally, an event is represented by a 600-dimensional vector. For a fixed number of voltage levels a two-step procedure of the position reconstruction was performed. First, the scintillator's volume was discretized and for each bin a high statistics set of reference 600-dimensional vectors was created. The objective of the second part of the procedure is to classify the new event to one of the given sets and hence determine the hit position. For more details about conducting the experiment of hit position reconstruction, the reader is referred to Ref.~\cite{NIM14LR}. 

\begin{figure}[h!]
	\centerline{\includegraphics[width=0.7\textwidth]{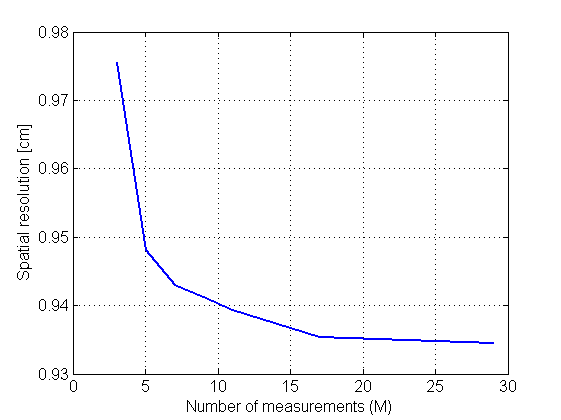}}
\caption{
The spatial resolution as a function of the acquired samples ($M$).
\label{spatial}
}
\end{figure}

We have conducted the test on the same data set and under the same conditions as described in Ref.~\cite{NIM14LR}, where the spatial resolution was reported to be equal to 1.05 cm ($\sigma$). The spatial resolutions derived from the recovered signals as a function of the number of samples $M$ included in the recovery process are shown in   
Fig.~\ref{spatial}.

In Fig.~\ref{spatial} only the region for small $M$, from 3 to 29, is shown, but it has to be stressed that the spatial resolution derived from the original raw-signal (300 samples) is equal to 0.933~cm ($\sigma$) and is almost the same as for $M$=29. For the most interesting case, with four voltage levels ($M$=7), the spatial resolution is slightly worse than for the fully sampled signal and is equal to 0.943~cm ($\sigma$). On the other hand, even in that case the spatial resolution is about 
0.1~cm better in comparison to the one evaluated based on signals in the voltage domain alone.  

\section{Conclusions}

In this paper a novel scheme of recovery of signals generated in plastic scintillator detectors in the J-PET scanner was introduced. The idea of signal recovery is based on the Tikhonov regularization theory, that uses the training data set of signals. In these studies we assumed that training signals come from a MVN distribution. The compact representation of these signals was provided by the PCA decomposition. 

One of the most important aspect of our work considers a statistical analysis of an error level of recovered signals. In this work a dependence of the signal recovery error on the number of samples taken in the voltage domain was determined. It has been proven that an average recovery error is approximately inversely proportional to the number of samples and inversely proportional to the decrease rate of variances in the covariance matrix. In the experimental section, the method was tested using signals registered by means of the single detection module of the J-PET detector. It was shown that the PCA basis offers high level of information compression and an accurate recovery may be achieved with just 8 samples for each signal waveform. It is worth noting that the developed recovery scheme is general and may be incorporated in any other investigation where a prior knowledge about the signals of interest may be utilized. 

In the experimental section we have demonstrated that using the recovered signals improves the hit-position reconstruction. In order to evaluate a position resolution of the J-PET scanner with fully recovered signals, we have incorporated the method for hit-position reconstruction, described in  Ref.~\cite{NIM14LR}. In the cited work, the spatial resolution evaluated on the same data set and under the same conditions, based on 8 samples in voltage domain, without a recovery of the waveform of the signal, was reported to be equal to about 1.05~cm ($\sigma$). Our experiment shows that the application of an information from four voltage levels to the recovery of the signal waveform can improve the spatial resolution to about 0.94~cm ($\sigma$). Moreover, the obtained result is only slightly worse than the one evaluated based on all 300 samples of the signals waveform. The spatial resolution calculated under these conditions is equal to about 0.93~cm ($\sigma$). It is very important information since, limiting the number of threshold levels in the electronic devices to four, leads to a reduction in the cost of the PET scanner.

Future work will address a development of the more advanced method to define the hit-position and event time for annihilation quanta in the J-PET detector based on the recovered information. We believe that, with fully recovered signals, there is still scope for improvement in the time and position resolution of the J-PET scanner.

\begin{appendices}

In order to prove the theorem we assume, for the sake of simplicity, that the matrix $A$ has normally distributed elements with zero means and $1/N$ variances. These values of the parameters of normal distribution ensure that the matrix $A$ is orthonormal. Hence, based on Eq.~(\ref{estS}), the matrix $S$ is given by:
\begin{equation}
	 S = \left(P^{-1} + \sigma^2 \frac{M}{N} \mathbb{1}\right)^{-1}. 
	 \label{estSaprox}
\end{equation}
The $\sigma_x^2$ is equal to the trace of the matrix $S$ and hence:
\begin{equation}
   \label{approxS}
  \begin{aligned}
    \sigma_x^2 &= \sum_{k=1}^{N} \frac{\sigma^2 N P_{k,k}}{\sigma^2 N + M P_{k,k}} 
    \\ 
     &= \frac{\sigma^2 N^2}{M} \left( 1 - \sigma^2 \sum_{k=1}^{N} \frac{1}{\sigma^2 N + M P_{k,k}} \right).
  \end{aligned}
\end{equation}
The sum in the last term in Eq.~(\ref{approxS}) may be approximated by a definite integral. In the following we will use for the calculations a basic, rectangle rule, and:
\begin{equation}
	 \sum_{k=1}^{N} \frac{1}{\sigma^2 N + M P_{k,k}} \approx \int_{1-h}^{N+h} \frac{1}{\sigma^2 N + M P(k)} dk = \textbf{I} \nonumber
\end{equation}
where $h = 1/2$. At the very beginning we assumed that the function $P(k)$ has the form: $P(k) = D \cdot e^{-\tau k}$ (see Eq.~(\ref{Pk})). We will perform the integration using the substitution $t =  e^{-\tau k}$. Without any significant loss of precision, we change the integration limits from $[1-h,~N+h]$ to $[0,~N]$. The calculations of the integral 
$\textbf{I}$ will be as follow:
\begin{align}
	\textbf{I} &= \int_{e^{-\tau N}}^{1}	\frac{1}{\left( \sigma^2 N + M D t \right) \tau t } dt \nonumber	\\
&= \int_{e^{-\tau N}}^{1}\frac{1}{\sigma^2 N \tau t}dt- \int_{e^{-\tau N}}^{1} \frac{MD}{\sigma^2 N\tau \left(\sigma^2 N+MDt \right) }dt
\nonumber		\\
&= \frac{1}{\sigma^2 N \tau}\left( \log(t)|_{e^{-\tau N}}^{1} - \log(\sigma^2 N + M D t)|_{e^{-\tau N}}^{1} \right)
\nonumber		\\
&\approx \frac{1}{\sigma^2 N \tau}\left(N \tau +\log \left(\frac{\sigma^2 N}{\sigma^2 N+MD} \right) \right) 
\nonumber
\end{align}
and thus
\begin{align}
\sigma_x^2 &\approx \frac{\sigma^2 N^2}{M} \left(1 - \sigma^2 \textbf{I} \right) 
\nonumber \\
&\approx \frac{\sigma^2 N}{M \tau} \cdot \log \left( \frac{\sigma^2 N+MD}{\sigma^2 N} \right).
\nonumber
\end{align}
\end{appendices}

\section{Acknowledgements}
We acknowledge technical and administrative support of T. Gucwa-Ry\'s, A. Heczko, M. Kajetanowicz, G. Konopka-Cupia\l, 
J. Majewski, W. Migda\l, A. Misiak, and the financial support by the Polish National Center for Development 
and Research through grant INNOTECH-K1/IN1/64/159174/NCBR/12, 
the Foundation for Polish Science through MPD programme, 
the EU and MSHE Grant No. POIG.02.03.00-161 00-013/09, 
Doctus - the Lesser Poland PhD Scholarship Fund, 
and Marian Smoluchowski Krak\'ow Research Consortium "Matter-Energy-Future". 
We are grateful to Prof. Colin Wilkin for correction of the manuscript.


%
\end{document}